\documentclass[aps,pre,twocolumn,showpacs,longbibliography,superscriptaddress,final,floatfix]{revtex4-1}
\usepackage{amsfonts} 
\usepackage{amsmath}
\usepackage{amssymb}
\usepackage{latexsym}
\usepackage{xcolor}
\usepackage{bm}
\usepackage{dcolumn}
\usepackage{epsfig}
\usepackage{graphicx}
\usepackage[breaklinks,colorlinks = true,linkcolor = red,urlcolor=blue,citecolor=red]{hyperref}
\usepackage{multirow}
\usepackage{makecell}
\usepackage[bbgreekl]{mathbbol}

\newcommand\ra{\rangle}
\newcommand\la{\langle}
\newcommand\nn{\nonumber}

\begin{document}

\title{Quantum Brownian Motion: Drude and Ohmic Baths as Continuum Limits of the Rubin Model}
 \author{Avijit Das}
 \affiliation{International Center for theoretical Sciences, Tata Institute of Fundamental Research, Bangalore-560089, India}
 \author{Abhishek Dhar}
 \affiliation{International Center for theoretical Sciences, Tata Institute of Fundamental Research, Bangalore-560089, India}
 \author{Ion Santra}
 \affiliation{Raman Research Institute, Bangalore-560080, India}
 \author{Urbashi Satpathi}
 \affiliation{International Center for theoretical Sciences, Tata Institute of Fundamental Research, Bangalore-560089, India}
 \author{Supurna Sinha}
 \affiliation{Raman Research Institute, Bangalore-560080, India}

\begin{abstract}
The motion of a free quantum particle in a thermal environment is usually described by the quantum Langevin equation, where the effect of the bath is encoded through a dissipative and a noise term, related to each other via the fluctuation dissipation theorem.  The quantum Langevin equation can be derived starting from a microscopic model of the thermal bath as an infinite collection of harmonic oscillators prepared in an initial equilibrium state. The spectral properties of the bath oscillators and their coupling to the particle determine the specific form of the dissipation and noise.
Here we investigate in detail the well-known Rubin bath model, which consists of a one-dimensional harmonic chain with the boundary bath particle  coupled to the Brownian  particle. We  show how in the limit of infinite bath bandwidth, we get the  Drude model and a second limit of infinite system-bath coupling gives the Ohmic model.  A detailed analysis of relevant  equilibrium correlation functions, such as  the mean squared displacement,  velocity auto-correlation functions, and the response function are presented, with the aim of  understanding of the various temporal regimes. In particular, we discuss the quantum to classical crossover time scales where the mean square displacement changes from a $\sim \ln t$ to a $\sim t$ dependence. We relate our study to recent work using linear response theory to understand quantum Brownian motion.
\end{abstract}

\maketitle

\section{Introduction}
\label{intro}

A good effective description for the motion of a classical Brownian particle in a thermal environment at temperature $T$ is given by the Langevin equation\cite{langevin1908}. Considering motion in one dimension this is given by 
\begin{align}\label{CLE}
M \dot{v}= -\gamma v +\eta(t)~, 
\end{align}
where $v=\dot{x}$ is the velocity of the particle, $x$ its position, $\gamma$ the dissipation constant and $\eta(t)$ is a Gaussian noise term with mean zero and correlations given by the fluctuation-dissipation relation $\langle \eta(t) \eta(t')\rangle = 2 \gamma k_B T \delta (t-t')$. Some of the most important properties of this effective dynamics are that the particle reaches thermal equilibrium with its velocity given by the Maxwell distribution with $\langle v \rangle=0$ and $\langle v^2 \rangle =k_B T/M$. On the other hand, the mean square displacement (MSD) shows diffusive growth at long times, $\Delta(t)=\langle [x(t)-x(0)]^2 \rangle = 2Dt$ (for $t \to \infty$), with a diffusion constant $D=k_B T /\gamma$.    

The quantum version of this equation was first written by Ford, Kac and Mazur \cite{Ford1965}. Unlike the classical case, where the Langevin equation  can be established using a purely phenomenological approach (see \cite{vanKampen2007}), the quantum case requires a microscopic modeling of the heat bath. The standard model for a heat bath is to treat it as an infinite collection of oscillators which is coupled to the system of interest, namely the Brownian particle. Eliminating the bath degrees, it can be shown that the effective dynamics of the particle is described by a quantum generalized Langevin equation, where the dissipation term has memory. A special choice of bath leads to the so-called Ohmic form \cite{weiss2012} of Eq.~\eqref{CLE}, with the noise correlations changed to the form
\begin{align}
\langle \eta(t) \eta(t')\rangle = \frac{ \gamma}{\pi}\int_0^\infty d\omega \hbar \omega [2 f(\omega,T)+1] \cos \omega(t-t')~,
\end{align} 
where $f(\omega,T)=[e^{\beta \hbar \omega}-1]^{-1}$ is the phonon distribution function. In particular we notice that in the quantum case, 
the noise is always correlated and there is no Markovian limit. Interestingly, even at zero temperature there is noise arising from 
quantum fluctuations and it has been shown that this leads to a logarithmic growth of the MSD with time : $\Delta_t \sim (\hbar/\gamma) \ln (t \gamma/M)$\cite{phillipson1974,hakim1985}. A pecularity of the quantum system is that the kinetic energy of the particle diverges \cite{Grabert1988}. This divergence arises due to the contribution of high frequency modes to the zero-point energy and  can be avoided by considering a finite bandwidth bath which leads to a damping term with memory. Since the original work of \cite{Ford1965}, quantum Brownian motion has been investigated using multiple approaches including quantum Langevin equations \cite{ford1988}, path integral methods \cite{caldeira1983,feynman1963}, equilibrium dynamical correlations\cite{balescu1975} and linear response theory\cite{balescu1975}. Other relevant references are \cite{hemmer1959,mazur1960,rubin1960,
turner1960I,turner1960II,rubin1961,
mazur1964,ullersma1966I,ullersma1966III,
zwanzig1973,presilla1996,presilla1997,hanggi2005,smith2008,lampo2019}.

In the present work, we discuss one of the simplest models of a quantum heat bath, the so-called Rubin bath \cite{rubin1960}. In general it corresponds to a bath with a dissipation kernel with long time memory, decaying as a power-law. However  we point out that as special limits it leads to the Ohmic bath (dissipation kernel is a delta function in time) and the Drude bath (dissipation kernel is exponentially decaying in time) \cite{weiss2012}. A different limiting procedure to obtain the Ohmic bath has been discussed in \cite{reybellet1999}. For the three bath models we discuss in detail the form of the MSD, as well as the velocity auto-correlation function and the response function,  all computed in the thermal equilibrium state. We try to understand interesting  physical aspects and highlight some of the qualitative differences. 
In recent years an approach based on linear response and fluctuation-dissipation theorem \cite{supurna1992,urbashi2017}  has been used to study Brownian motion at zero temperature. 
We point out here that this  approach is exact for the case of the Rubin model of bath. 

We note that quantum Langevin equations were 
discussed in the most general setting in \cite{ford1988} where 
the authors first discuss these equations without resorting to a 
microscopic model of the bath and state the necessary conditions on the 
memory kernel that appears in the dissipation term. The paper then discusses 
how the independent oscillator model of a heat bath can be used to derive the 
quantum Langevin equation of the most general form. It is also pointed out that 
various bath models that were studied earlier are special cases of the independent 
oscillator (IO) model. Some of the linear coupling models that do not fall in this class 
do not satisfy the so-called positivity condition and are unphysical. The choice  of 
frequencies of the bath oscillators and their coupling to the test particle fixes the 
memory kernel and the complete dynamics. However, the work of \cite{ford1988}  discusses the general 
formalism but does not investigate the interesting physical properties that are observed for specific 
choices of bath models. The present paper explores precisely these aspects for a particular choice of bath, 
namely the Rubin model, which naturally falls in the class of IO models. For completeness here we first briefly 
outline the steps which leads from the Rubin model to the IO model, and then to the quantum Langevin equation. 
As mentioned above, our other important contribution is to point out that special limits of the Rubin model lead 
to two physically relevant bath models, namely the Ohmic and Drude models.  The corresponding dissipation kernels 
for these models are simple and so are widely used, and a natural question is whether the use of these models as 
approximate descriptions of the original Rubin model preserves some of the observable properties. Our comparative 
study of the properties of various physical obsrvables for the three models throws light on this question.

This paper is organized as follows. In Sec.~(\ref{model}) we introduce the Hamiltonian and derive the generalized Langevin equation for the system by integrating out the bath degrees of freedom. We have also discussed a continuum limit of the model and shown that the conventional and simpler models of the bath, Drude and Ohmic, emerge. In Sec.~(\ref{correlations}) we define the relevant correlation functions - : mean square displacements, velocity autocorrelation functions, and the response functions. In Sec.~(\ref{comparison}) we compute these correlations and compare the different models in detail. We end the paper with a few concluding remarks in Sec.~(\ref{Summary}).

\section{Hamiltonian and derivation of the generalised Langevin equation}\label{model}
\begin{figure}
    \centering
    \includegraphics[scale=0.2]{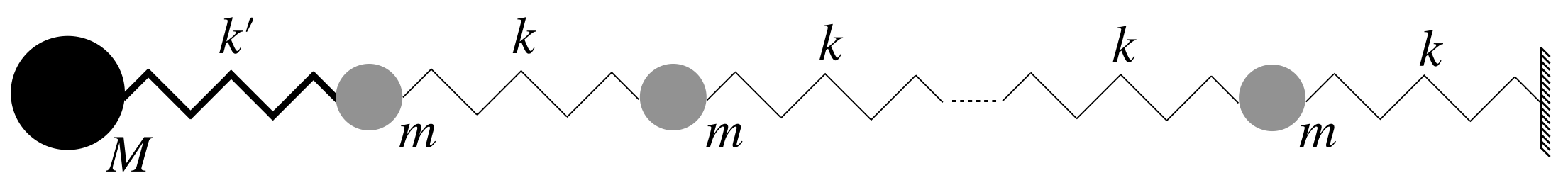}
    \caption{Setup of the problem.}
    \label{fig1}
\end{figure}

Our set-up consisting of a single particle  coupled to the  Rubin bath 
is schematically shown in Fig.~(\ref{fig1}).  
We consider a particle of mass $ M $ with position and momentum  operators 
specified by $x$ and $p$ respectively, while the bath consists   of $N$  particles of mass $m$ and 
position and momentum operators given by $\{x_j,p_j\}$, $j=1,2,\ldots N$  that are coupled by harmonic springs of stiffness $k$. 
The Hamiltonian of the coupled system and bath is given by
\begin{align}\label{eq1}
\mathcal{H} = \frac{p^2}{2M} &+\frac{k'}{2}(x-x_1)^2 \nonumber\\
 &+\sum_{n=1}^{N}\frac{p_n^2}{2m}+\frac{k}{2}\sum_{n=1}^{N}(x_n -x_{n+1)})^2,
\end{align}
where we consider the right boundary to be fixed $x_{N+1}=0$.   Even though our test particle $(x,p)$ is tied to the bath, we will see that in the limit $N \to \infty$, the effective motion corresponds to that of a free particle. For our analysis, it is convenient to write the above Hamiltonian in the following form:
\begin{align}
&\mathcal{H}  = \mathcal{H}_S + \mathcal{H}_B + \mathcal{H}_{SB}~, \label{bathH} \\
&{\rm where}~~\mathcal{H}_S=\frac{p^2}{2M} + \frac{k'}{2} x^2~,~~ \mathcal{H}_{SB}=-k' x x_1~, \nn \\
& \mathcal{H}_B = \sum_{n=1}^{N} \frac{p_n^2}{2m} + \frac{k'}{2}x_1^2+\frac{k}{2}\sum_{n=1}^{N-1}(x_n -x_{n+1)})^2+\frac{k}{2}x^2_{N+1}~. \nn
\end{align}
The bath Hamiltonian can be written in the compact form $\mathcal{H}_B =  {\bf p}^Tm^{-1} {\bf p}/2+{\bf x}^T \phi {\bf x}/2$, 
where ${\bf x}=(x_1,x_2,\ldots,x_N)$ and ${\bf p}=(p_1,p_2,\ldots,p_N)$ and  $\phi$ is the force matrix. 
Let us consider a linear transformation ${\bf X}=m^{1/2}U {\bf x}$ and ${\bf P}=m^{1/2}U {\bf p}$ where $U$ is an orthogonal 
transformation which diagonalizes the force matrix, i.e, $U \phi U^T= m \Omega^2$, where $\Omega^2$ is the 
diagonal matrix with elements given by the normal mode frequencies of the bath $\Omega^2=\{\Omega_s^2\}$, with $s=1,2,\ldots,N$. Note that the  column vector formed by the matrix 
elements $U_{si}$ gives the normal mode eigenfunction corresponding to the eigenvalue $\Omega_s^2$. Using the normal mode coordinates $X_s$ and momenta $P_s$ the system-bath coupling and the bath Hamiltonian  can be written as
\begin{align}
\mathcal{H}_{SB}&=- k' x x_1 = -k' \sum_{s=1}^N C_s x X_s,~ {\rm where}~ C_s=m^{-1/2}U_{s1} \nn \\
\mathcal{H}_B&=\sum_{s=1}^{N} \frac{P_s^2}{2}+ \frac{\Omega_s^2 X_s^2}{2} ~.
\end{align}
To derive the effective Langevin equations  for the system, one starts by writing  the Heisenberg equations of motion of the system and the bath degrees of freedom given by
\begin{align}
 M\ddot{x} &= -k' x + k' \sum_{s=1}^N C_s X_s, \label{eqmS}\\
\ddot{X}_s &= -\Omega^2_s X_s + k' C_s x~,~~s=1,2,\dots,N. \label{eqmB}
\end{align}
The bath equations of motion Eq.~\eqref{eqmB} can be solved formally, assuming initial conditions $\{X_s(t_0),P_s(t_0)\}$ that are chosen, at time $t_0$, from the Boltzmann distribution $ e^{-\beta \mathcal{H}_B}/\rm{Tr}\left[e^{-\beta \mathcal{H}_B}\right]$ at temperature $T=(k_B \beta)^{-1}$. This gives
\begin{align}
X_s(t)&=\cos \left\{\Omega_s(t-t_0)\right\} X_s(t_0) + \frac{\sin \left\{\Omega_s(t-t_0)\right\} }{\Omega_s}P_s(t_0) \nonumber \\
&+k' C_s \int_{t_0}^t dt'  \frac{\sin \left\{ \Omega_s(t-t')\right\} }{\Omega_s} x(t')~.
\end{align}
Plugging this into the  equation of motion of the system we get 
\begin{equation}
 M\ddot{x} = -k'x +\int_{t_0}^t dt'\Sigma (t-t') x(t') + \eta(t)~, \label{eqGLE}
\end{equation}
 where
\begin{align} \label{DisiNoise}
 \Sigma(t)&=k'^2 \sum_{s=1}^{N} C_{s}^2 \frac{\sin(\Omega_s t)}{\Omega_s}, \nn \\
 \eta(t) &=k'\sum_{s=1}^{N} C_{s} \Big[ \cos \left\{\Omega_s(t-t_0)\right\} X_s(t_0) \nn\\  &\hspace{50pt}+ \frac{\sin\left\{\Omega_s(t-t_0)\right\}}{\Omega_s} P_s(t_0) \Big].
\end{align}
The form in Eq.~\eqref{eqGLE} is in the form of a  generalized Langevin equation, with $ \Sigma(t)$ as the memory kernel and $\eta(t)$ as the random force term. The information about the baths is completely contained in these two terms. To get a valid bath it is necessary to take the limit $N \to \infty$ since otherwise we would hit  {Poincar\'e} recurrences \cite{weiss2012}.
 Indeed, apparent dissipation arises in a Hamiltonian system because of the flow of energy into the infinite degrees of freedom of the bath. 
Secondly since we are interested in equilibrium properties we next take the 
 $t_0 \to -\infty$ limit (after the $N\to \infty$ limit). This ensures that at any finite time the Brownian particle has reached thermal equilibrium. Mathematically the $t_0 \to -\infty$ limit allows us to use Fourier transforms and equilibrium correlations can be readily computed. 
It is instructive to write the above equations in the usual form of Langevin equations where the dissipation term involves the velocity rather than the positional degree of freedom. For this we define the dissipation kernel 
\begin{align}
\gamma(t)=k'^2 \sum_{s=1}^{N} C_{s}^2 \frac{\cos(\Omega_s t)}{\Omega^2_s},
\end{align}
 so that,
\begin{equation}\label{eq_sigma_gamma}
\Sigma(t)= -\frac{d\gamma(t)}{dt}~.
\end{equation}
We plug this into Eq.~\eqref{eqGLE} and perform an integration by parts.  Then, using  the  identities $\gamma(0)= k'^2 \sum_{s=1}^N \frac{C_s^2}{\Omega_s^2} = k'^2 \left[ \phi^{-1} \right]_{11} = k'$ and $\gamma(\infty)= 0$, which can be proved in the $N\to \infty$ limit (for a reasonable choice of bath properties which are indeed satisfied by the baths we have considered here) and setting $t_0 \rightarrow -\infty$, we get
\begin{align}
 M\ddot{x} = -\int_{-\infty}^t dt'\gamma (t-t') \dot{x}(t') + \eta(t)~, \label{eqGLEgamma}
\end{align}
 where we now see that the pinning potential does not appear, which is what  one  would like for a free  particle.
We will now compute the bath properties in the $N \to \infty$ limit.  It is useful to define the  bath spectral functions 
\begin{align}
&\Sigma^+(\omega)=\int_0^\infty dt \Sigma(t) e^{i \omega t}={k'}^2\sum_s \frac{C_s^2}{-(\omega+i \epsilon)^2 +\Omega_s^2} \nn \\ 
&\Gamma(\omega)=\rm{Im}\left[\Sigma^+(\omega)\right]={k'}^2 \sum_s \frac{ \pi C_s^2}{2 \omega}\left[\delta (\omega-\Omega_s) + \delta (\omega+\Omega_s)\right].
\end{align}
The statistical properties of the noise term can be obtained using the fact that  at $t=t_0$ the bath is in thermal equilibrium at temperature $T$. 
Thus we find that $\langle \eta(t) \rangle = 0$ while the noise correlations are easiest to state in the Fourier domain. Defining $\tilde{\eta}(\omega)=
\int_{-\infty}^\infty dt \eta(t) e^{i \omega t}$, we find \cite{weiss2012} 
\begin{align}
\langle \tilde{\eta}(\omega) \tilde{\eta}(\omega')\rangle &= 4\hbar\pi~ \Gamma(\omega) [f(\omega,T)+1] ~\delta(\omega+\omega')~, \label{noisecorr} 
\end{align}
where $f(\omega,T)=[e^{\beta \hbar \omega}-1]^{-1}$ is the phonon distribution function. To compute  $\Sigma^+(\omega)$, we note that it is precisely given by $k^{'2}~g^+_{1,1}$ where $g^+=[-m (\omega +i \epsilon)^2 +\phi]^{-1}$ is the phonon Green's function of the heat bath and $g^+_{11}$ refers to its diagonal element at site $n=1$, corresponding to the particle that is coupled to the system. The computation of $g^+(\omega)$ becomes a bit involved because of the presence of the ``impurity'' term in the bath Hamiltonian $\mathcal{H}_B$ in Eq.~(\ref{bathH}). However, this can still be obtained explicitly and one eventually obtains \cite{das2012} 
\begin{eqnarray}\label{eq16}
\Sigma^{+}(\omega)&=& k'^{2}\frac{e^{iq}}{k+(k'-k)e^{iq}},
\end{eqnarray}
where $q$ is given by the solution of the dispersion $\omega^2 = (2k/m) (1-\cos q)$. In the frequency range $|\omega| \leq 2 \sqrt{k/m}$, we get real values for $q$ and then we have
\begin{align}\label{eq17}
\Gamma(\omega)=\frac{k'^{2} k \sin q}{|k'-k+k e^{-iq}|^2}= \frac{k'^{2}}{k}\frac{\omega \sqrt{\frac{m}{k}} \sqrt{1 - \frac{m \omega^2}{4 k}}}{(\frac{k'}{k})^2+ (1-\frac{k'}{k}) \frac{m \omega^2}{k}},
\end{align}
while for $|\omega| > 2 \sqrt{k/m}$, we get $\Gamma(\omega)=0$. The real part of $\Sigma^+(\omega)$ is the following:
\begin{align}\label{eq18}
\mathrm{Re} \left[ \Sigma^+(\omega) \right] = 
\begin{cases}
\frac{k'^{2}}{k}\frac{\frac{k'}{k} - \frac{m\omega^2}{2k}}{\left(\frac{k'}{k}\right)^2+ \left(1-\frac{k'}{k}\right) \frac{m \omega^2}{k}} \hspace{22pt}; & |\omega| \leq 2\sqrt{\frac{k}{m}}\vspace{12pt}\\
\frac{k'^{2}}{k}\frac{\frac{k'}{k} - \frac{m\omega^2}{2k} + \frac{m\omega^2}{2k} \sqrt{1-\frac{4 k}{m\omega^2}}}{\left(\frac{k'}{k}\right)^2+ \left(1-\frac{k'}{k}\right) \frac{m \omega^2}{k}} ~;& |\omega| > 2\sqrt{\frac{k}{m}}.
\end{cases}
\end{align}
Note that $\mathrm{Re} \left[ \Sigma^+(\omega) \right]$ is even with respect to $\omega$ whereas $\Gamma(\omega) = \mathrm{Im} \left[ \Sigma^+(\omega) \right]$ is an odd function of $\omega$. $\Sigma^+(\omega)$ decays to zero for $|\omega|\rightarrow\infty$ which is necessary for its Fourier transform $\Sigma(t)$ to exist. These expressions of $\Sigma^+(\omega)$ become particularly simple for the case $k=k'$. Finally we note that $\tilde{\gamma}(\omega)=\int_0^\infty dt \gamma(t) e^{i \omega t}$ is given by,
\begin{align}\label{eq19}
i \omega \tilde{\gamma}(\omega)=\Sigma^+(\omega)-k'~.
\end{align}

\emph{Continuum string limit}: An interesting special case is to consider the limit corresponding to the bath being a continuous string. This has been discussed earlier in \cite{reybellet1999} but in a somewhat different setting. We introduce a lattice spacing $a$ and define the mass density $\sigma=m/a$, Young's modulus $E=ka$. The lattice parameter can be introduced in Eq.~\eqref{eq17} and \eqref{eq18} in a consistent way by substituting $m = \sigma a$, $k = E/a$, etc. The continuum limit is obtained by taking $a \to 0$, $m \to 0$ and $k \to \infty$ while keeping $E$ and $\sigma$ constant. This then gives
\begin{align}\label{eq_20}
\Gamma(\omega)&= \frac{ \gamma_0 \omega}{1+\omega^2 \tau^2}~,~~
\tilde{\gamma}(\omega)=\frac{\gamma_0}{1-i \omega \tau}, \nn\\
{\rm where}~~\gamma_0&=(\sigma E)^{1/2},~~\tau=\gamma_0/k'~. 
\end{align}
This corresponds to the so called \emph{Drude} model of the bath, corresponding to a dissipation kernel $\gamma(t)=(\gamma_0/\tau) e^{-t/\tau}$. Taking the strong coupling limit $k' \to \infty$, so that $\tau \to 0$, gives us the \emph{Ohmic} bath model with
\begin{align}\label{eq_21}
\Gamma(\omega)= \gamma_0 \omega,~\tilde{\gamma}(\omega)=\gamma_0~,
\end{align}
 which gives us a memory-less dissipation kernel $\gamma(t)=\gamma_0 \delta(t)$. We note that the presence of the phonon distribution function $f(\omega,T)$ in the quantum system ensures that the noise in Eq.\eqref{noisecorr} is still correlated and has memory. However, in the high temperature limit, $\beta \hbar \omega \rightarrow 0$, we achieve the strictly Markovian limit $\langle \eta(t) \eta(t') \rangle = 2 \gamma_0 k_B T \delta(t-t')$. The authors in \cite{reybellet1999}  obtained the Ohmic bath  starting from a continuum field description of the bath and using a different limiting procedure. 

In the next section we discuss the behavior of various physical observables for the quantum Brownian particle that are obtained from the Rubin model and its  limiting forms.

\section{Mean Square Displacement, Velocity Autocorrelation Function  and  Response Function}\label{correlations}

In the long time limit the particle reaches the equilibrium  state 
and  we focus on properties in this state such as the mean square displacement, the velocity autocorrelation function and response functions.  
The mean square displacement and the velocity autocorrelation function are defined as
\begin{align}
\Delta(t) &=\big\langle \big( x(t)-x(0) \big)^2 \big\rangle, ~C(t)= \frac{1}{2}\la \{v(t), v(0)\} \ra, \nn
\end{align}
where $\{\ldots \}$ denotes the anticommutator. 
The response function $R(t)$ and velocity response function $\bar{R}(t)$ are defined through the following equations for the average displacement and average velocity in the presence of a driving force $f(t)$.
\begin{align}
\la \Delta x(t) \ra &:=  \la x(t) \ra_f -  \la x \ra_{f=0} = \int_{-\infty}^t dt' R(t-t') f(t')\\
\la v(t) \ra &= \int_{-\infty}^t dt' \bar{R}(t-t') f(t')~,
\end{align}
where $\la \cdots \ra_f$ is the expectation value in the presence of the force and $\la \cdots \ra_{f=0}$  in the absence of it. By definition $\bar{R}(t)=\dot{R}(t)$. All these three quantities can be obtained through the Fourier transform solution of Eq.~\eqref{eqGLE} (after taking the limits $N \rightarrow \infty$ and $t_0 \rightarrow -\infty$) and \eqref{eqGLEgamma}. The transform $\tilde{x}(\omega) = \int_{-\infty}^\infty dt x(t) e^{i \omega t} $ is given by  
\begin{align}
\tilde{x}(\omega)&=G(\omega) \tilde{\eta}(\omega),~~{\rm where} \\ 
G(\omega)&=\frac{1}{-M\omega^{2}+k'-\Sigma^{+}(\omega)} = 
\frac{1}{-M \omega^2 -i \omega \tilde{\gamma}(\omega)}~.\label{eq25}
\end{align}
Using this and the noise properties leads immediately to
\begin{align}
\Delta(t)& =2 \langle x^2(0)\rangle -\langle\left\lbrace x(t),x(0) \right\rbrace\rangle \\
&\hspace{-15pt}=\frac{\hbar}{\pi} \int_{-\infty}^{\infty}d\omega \coth (\beta \hbar \omega/2)  \Gamma(\omega) G(\omega) G(-\omega) \left(1-e^{-i\omega t}\right)~ \nn \\
&\hspace{-15pt}=\frac{2 \hbar}{\pi} \int_{0}^{\infty}d\omega \coth (\beta \hbar \omega/2)  \Gamma(\omega) G(\omega) G(-\omega) \left(1-\cos{\omega t}\right)~ \nn \\
&\hspace{-15pt}= \frac{2\hbar}{\pi} \int_{0}^{\infty}d\omega \coth (\beta \hbar \omega/2){\rm Im} \left[ G(\omega) \right] \left(1-\cos{\omega t}\right), \label{eq_delta_t}
\end{align}
where we used the Green's function identity  $ \Gamma(\omega)G(\omega) G(-\omega) = [G(\omega)-G(-\omega)]/(2i)$.

\begin{figure*}
    \centering
    \includegraphics[width=\textwidth,height=5cm]{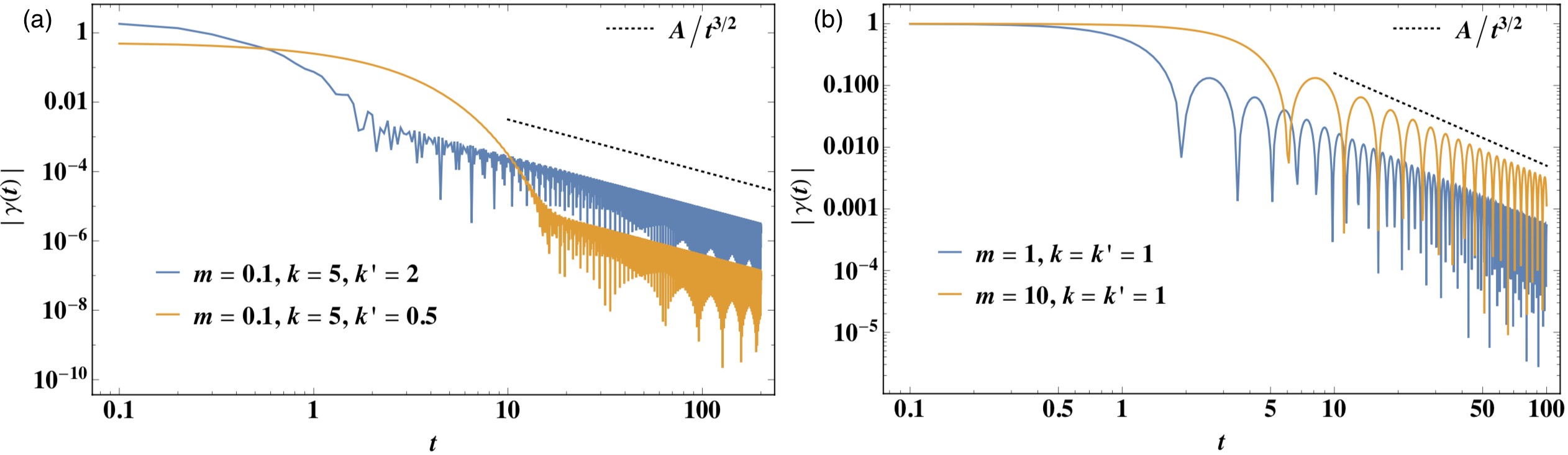}
   \caption{\textbf{(Rubin model)} Log-Log plot of $|\gamma(t)|$ for a set of values of 
$m,k,k'$ to show that $\gamma(t)$ initially decays fast and then as a power law 
$\sim t^{-3/2}$. We propose that the crossover time, $t^*$, can be estimated from the location of the branch point of 
$\tilde{\gamma}(\omega)$: $t^* \propto \sqrt{\frac{m}{k'}\left( \frac{k}{k'} -1 \right)}$ 
when $k>k'$ and $t^* \propto \sqrt{\frac{m}{k}}$ for $k\leq k'$. 
(a) $(k>k')$; if we decrease just $k'$ by a factor of $4$ 
keeping other parameters fixed, $t^*$ increases $4$ times. 
(b) $(k=k')$; $m$ is increased by $10$ times, which results in a shift of $t^*$ by a factor of $\sqrt{10}$. These observations support our claim about the crossover time.}
    \label{fig2}
\end{figure*}

\begin{figure*}
    \centering
    \includegraphics[width=\textwidth,height=5cm]{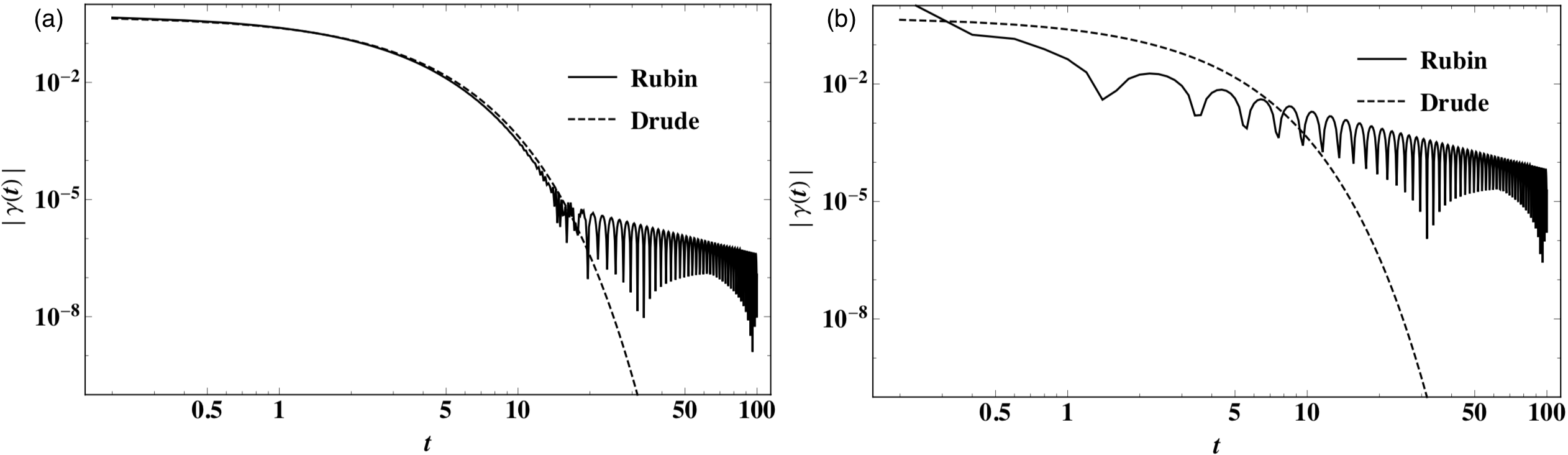}
   \caption{\textbf{Comparison of $\bm{\gamma(t)}$ between the Rubin and Drude models:} (a) Weak coupling case $\bm{k' = 0.5}$; (b) Strong coupling case with $\bm{k' = 4.0}$. Other parameters are taken as $M = 1, m = 0.1, k = 5$. This data supports the fact that the Drude approximation of the Rubin bath is good when $k$ is large but $k'$ is not. If both $k$ and $k'$ are made large, the Ohmic approximation is better than the Drude.}
\label{gammacomp}
\end{figure*}

The velocity auto-correlation function can be obtained from $\Delta(t)$ as 
\begin{align}
C(t)&= \frac{1}{2}\frac{d^2 \Delta(t)}{dt^2} \label{eq_ct_def}  \\
&=\frac{\hbar}{\pi} \int_{0}^{\infty}d\omega \coth (\beta \hbar \omega/2)  \Gamma(\omega) G(\omega) G(-\omega) \omega^2 \cos{\omega t}\nn\\
&=\frac{\hbar}{\pi} \int_{0}^{\infty}d\omega \coth (\beta \hbar \omega/2)   {\rm Im} \left[ G(\omega) \right] \omega^2 \cos{\omega t}~.
\end{align}
The velocity response function is given by
\begin{align}
\bar{R}(t)&=\frac{1}{2 \pi} \int_{-\infty}^\infty d \omega \frac{ e^{-i \omega t}}{-i\omega M+\tilde{\gamma}(\omega)}\\
&= \frac{1}{2 \pi} \int_{-\infty}^\infty d \omega (-i\omega) G(\omega) e^{-i \omega t}~\label{eq_c_t}.
\end{align}
Whereas the relation $\bar{R}(t)=\dot{R}(t)$ gives us an expression of the position response function,
\begin{align}\label{eq_R}
&R(t)=\int_0^tdt' \bar{R}(t') = \frac{1}{2 \pi} \int_{-\infty}^\infty d \omega ~ G(\omega) (e^{-i \omega t}-1) \\
&\hspace{-10pt}= \frac{1}{\pi} \int_0^\infty d\omega \Big( \mathrm{Re}[G(\omega)][\cos(\omega t) -1] + \mathrm{Im}[G(\omega)] \sin(\omega t) \Big),
\end{align}
where we have used the symmetry properties of $G(\omega)$: $\mathrm{Re}[G(-\omega)] = \mathrm{Re}[G(\omega)]$ and $\mathrm{Im}[G(-\omega)] = - \mathrm{Im}[G(\omega)]$. On the other hand, the positional correlation function is given by
 \begin{align}
\frac{1}{i \hbar}\la [x(t),x(0)]\ra &= \frac{1}{\pi i} \int_{-\infty}^\infty d \omega \Gamma(\omega) G(\omega)  G(-\omega) e^{-i \omega t}~ \nn \\
&= \frac{-1}{2 \pi} \int_{-\infty}^\infty d \omega [G(\omega) -G(-\omega)]   e^{-i \omega t}~ \nn \\
&= \frac{-1}{\pi} \int_{-\infty}^\infty d \omega {\rm Im}[G(\omega)]\sin(\omega t)~.
\end{align}
Using the Kramer's Kronig identity, $\int_{-\infty}^\infty d \omega ~{\rm Im}[G(\omega)] \sin (\omega t) = \int_{-\infty}^\infty d \omega ~ {\rm Re}[G(\omega)] [\cos (\omega t)-1]$, we verify explicitly that the linear response formula
\begin{align}
R(t)=-\frac{1}{i \hbar} \la [x(t),x(0)]\ra,
\end{align}
holds exactly. This is expected since the dynamics of system and bath is completely linear.

\section{Comparison of the forms of $\gamma(t), \Delta(t)$ and $C(t)$ for the three models}\label{comparison}

\begin{figure*}
    \centering
    \includegraphics[width=\textwidth,height=5cm]{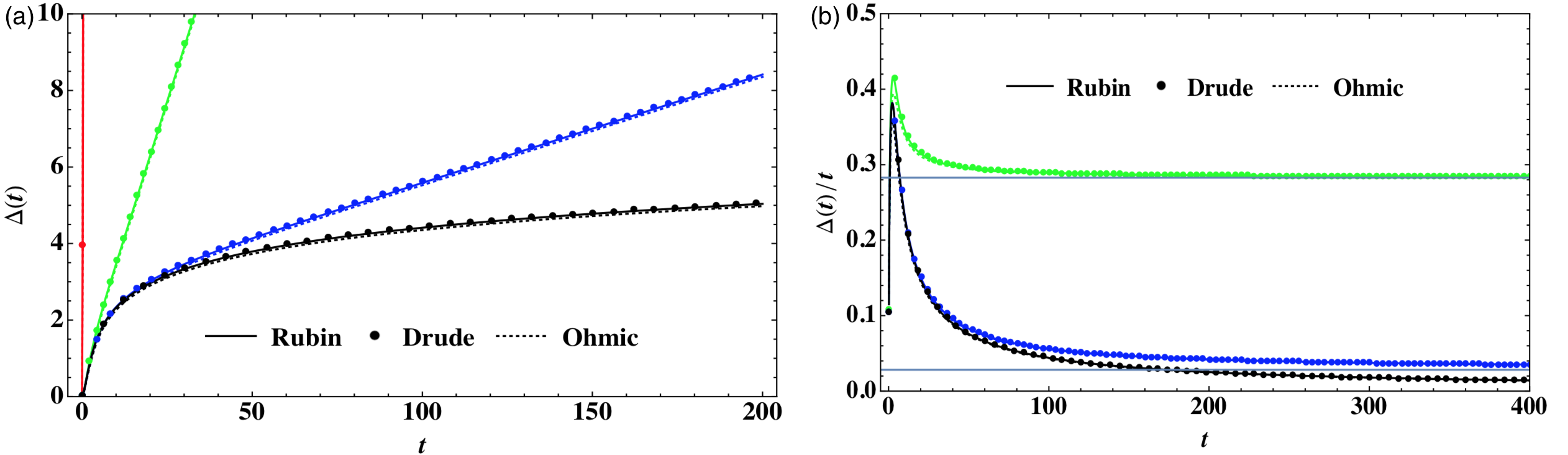}
   \caption{(a) Comparison of $\Delta(t)$ between the Rubin, Drude and Ohmic models. T
he $\beta$ values from the top are $0.01, 10, 100$ and $\infty$ (from above). 
(b) For the $\beta$ values $10, 100, \infty$ (from above) we plot $\Delta(t)/t$. 
Other parameters are taken as $M = 1, m = 0.1, k = 5, \bm{k' = 4.0}$. 
This figure is the counterpart of Fig.~(\ref{deltacomp}) with $k$ and $k'$ both large. 
The saturation values ($0.283$ and $0.0283$) are indicated in the figure. We see a agreement between the three models compared to 
Fig.~(\ref{deltacomp}).}
\label{fig6}
\end{figure*}

\begin{figure*}
    \centering
    \includegraphics[width=\textwidth,height=5cm]{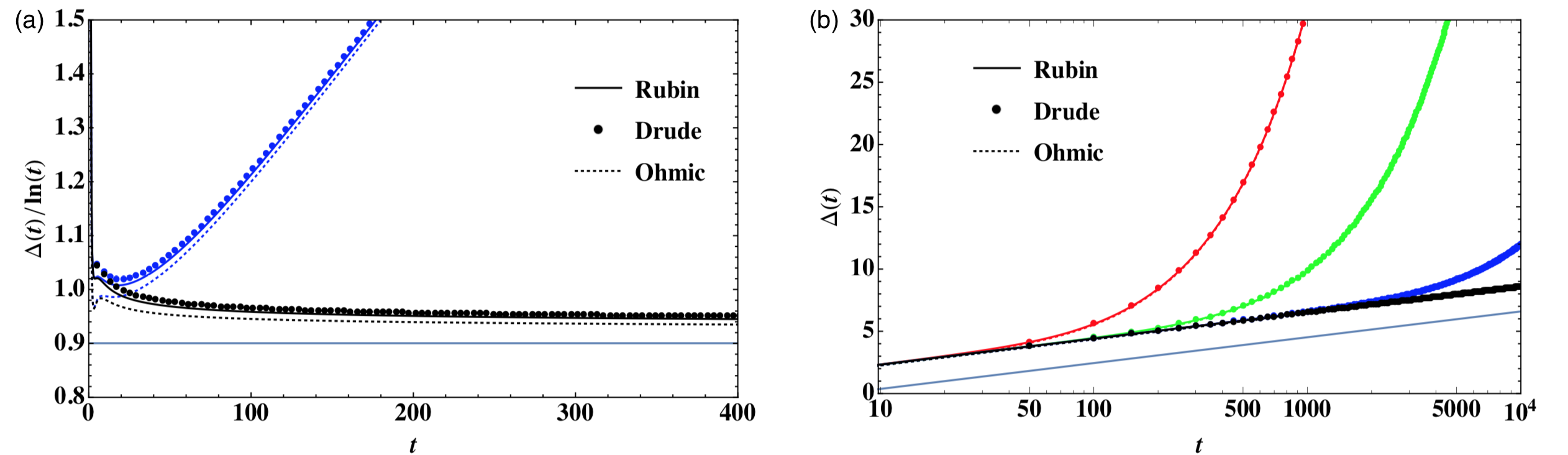}
   \caption{(a) Comparison of $\Delta(t)/\ln (t)$ between the Rubin, Drude and Ohmic models for $\beta=10,\infty$ (from above) in linear scale. 
(b) $\Delta(t)$ in log-linear scale for $\beta = 100, 500, 5000, \infty$ (from above). 
Other parameters were taken as $M = 1, m = 0.1, k = 5, \bm{k' = 4.0}$. Note the match between different models, as $k$ and $k'$ both are large 
in contrast to  Fig.~(\ref{deltacompLT}). 
From Eq.~\eqref{eq_43} the prefactor of $\ln(t)$ is $2\hbar/\pi\gamma_0 = 0.9$ which has been indicated both in (a) and (b).}
\label{fig7}
\end{figure*}

\begin{figure*}
    \centering
    \includegraphics[width=\textwidth,height=5cm]{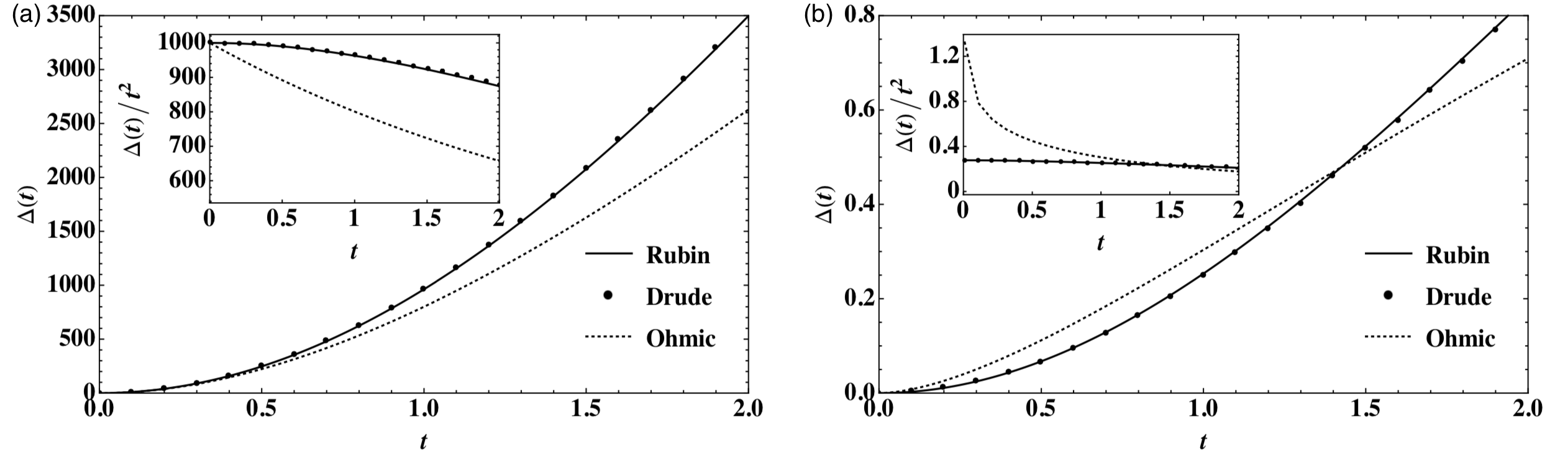}
   \caption{Comparison of $\Delta(t)$ between the Rubin, Drude and Ohmic models at short times. 
(a) $\beta = 0.001$. We can see that at high temperatures $\Delta(t)$ behaves as $\sim t^2$ for all three models. 
The correction to the $t^2$ behavior for the Ohmic case is $~t^3$, which is evident from the inset. 
(b) $\beta=100$. At low and finite temperatures, the short time behavior is $\sim -t^2 \ln(t)$ for the Ohmic bath 
whereas it is $\sim t^2$ for the Rubin and Drude baths. Other parameters are $M = 1, m = 0.1, k = 5, k' = 0.5$. 
Note that the log divergence of $C(0)$ for the Ohmic case is present at any finite temperature and diminishes for $\beta \hbar$ is equal to zero, 
which is hard to achieve numerically. Thus in the data presented in (a) for the Ohmic case, we have taken the classical limit first and then performed the integral.}
\label{ShortTimeDelta}
\end{figure*}

\subsection{Form of $\gamma(t)$}

{\bf Rubin model}: In this case one can obtain the expression of $\tilde{\gamma}(\omega)$ using Eq.~\eqref{eq16} and Eq.~\eqref{eq19}:

\begin{align}\label{eq_re_gamma_Rubin_kneq_k1}
\mathrm{Re}\left[ \tilde{\gamma}(\omega) \right] = 
\frac{k'^2}{k} \sqrt{\frac{m}{k}} \frac{ \sqrt{1-\frac{m \omega ^2}{4 k}}}{ \left(\frac{k'}{k}\right)^2+\left(1-\frac{k'}{k}\right) \left(\frac{m \omega ^2}{k}\right)}
\end{align}
for $|\omega| \leq 2\sqrt{\frac{k}{m}}$ and $\mathrm{Re}\left[ \tilde{\gamma}(\omega) \right]=0$ for $|\omega| > 2\sqrt{\frac{k}{m}}$.
\begin{align}\label{eq_im_gamma_Rubin_kneq_k1}
\mathrm{Im}\left[ \tilde{\gamma}(\omega) \right] =
m \omega \frac{ \frac{k'}{k} - \frac{1}{2}\left(\frac{k'}{k}\right)^2 }{ \left(\frac{k'}{k}\right)^2+\left(1-\frac{k'}{k}\right) \left(\frac{m \omega ^2}{k}\right)}
\end{align}
for $|\omega| \leq 2\sqrt{\frac{k}{m}}$ and
\begin{align}
\mathrm{Im}\left[ \tilde{\gamma}(\omega) \right] = m \omega \frac{ \frac{k'}{k} - \frac{1}{2}\left(\frac{k'}{k}\right)^2 \left[ 1+ \sqrt{1-\frac{4 k}{m \omega ^2}} \right] }{ \left(\frac{k'}{k}\right)^2+\left(1-\frac{k'}{k}\right) \left(\frac{m \omega ^2}{k}\right)}
\end{align}
for $|\omega| > 2\sqrt{\frac{k}{m}}$.


Note that $\mathrm{Re}\left[ \tilde{\gamma}(\omega) \right]$ is odd function of $\omega$ while $\mathrm{Im}\left[\tilde{\gamma}(\omega) \right]$ is even. This property is common for various response functions in physical systems. For the special case $k=k'$, it is possible to evaluate $\gamma(t) = \frac{1}{2\pi}\int_{-\infty}^\infty d\omega~ \tilde{\gamma}(\omega) e^{-i\omega t}$ to obtain
\begin{align}
\gamma(t) &= \frac{\sqrt{k m} J_1\left(2 \sqrt{\frac{k}{m}} t\right)}{t}~,
\end{align}
where $J_n$ is the Bessel function of 1st kind.
 Since $J_n(x) \sim \sqrt{\frac{2}{\pi x}} \cos\left[ x - ( n + 1/2 ) \frac{\pi}{2} \right]$ at large $x$,  we get the leading order asymptotic behavior $\gamma(t) \sim t^{-3/2}$. 
This leading asymptotic form can be seen as arising from the branch point at $\omega =2 \sqrt{k/m}$ in the integrand in Eq.~\eqref{eq_re_gamma_Rubin_kneq_k1}.  
For the general case, $k \neq k'$, we note that the integrand has additional poles at $\omega=k'/\sqrt{m(k'-k)}$. For $k>k'$, this is imaginary and gives rise to an exponentially decaying part in $\gamma(t)$. Thus we expect that  for $k >k'$, $\gamma(t)$ should initially have a fast exponential decay $\sim e^{-w_p t}$,where $\omega_p=k'/\sqrt{m(k-k')}$. After a time scale $t_c \approx 2 \pi/\omega_p$, this is followed by a $\sim t^{-3/2}$ decay.  
This feature is clearly seen in the  numerical evaluation of $\gamma(t)$ is presented in Fig.~(\ref{fig2}) for two different parameter sets. In \cite{rajdeep2018} the authors have addressed this question of crossover timescales in a similar system heuristically.\\

{\bf Drude bath and Ohmic bath limits}: From Eq.~\eqref{eq_20} one obtains $\gamma(t) = \frac{\gamma_0}{\tau}e^{-t/\tau}$ and $ \Sigma(t)= \frac{\gamma_0}{\tau^2}e^{-t/\tau}$. Ohmic bath is obtained simply taking the limit $\tau \rightarrow 0$ and gives $\gamma(t) = \gamma_0 \delta(t)$.  

In Fig.~\eqref{gammacomp} we show a comparison of the forms of $\gamma(t)$ obtained from the Rubin and Drude models. As expected we see that for the weak-coupling case ($k'=0.5$), we expect an exponential decaying regime for the Rubin model over the time-scale $t_c \approx 2 \pi/\omega_p \approx 25.3$  and here we see agreement with the Drude model. On the other hand, when $ k'=4.0$, we see that $t_c \approx 1.5$ and correspondingly one finds that there is no regime where the Drude approximation is good. 

We  next explore the question on how well the behavior of other physical observables  such as $\Delta (t)$ and $C(t)$ are reproduced by the Drude and Ohmic approximations.

\subsection{Form of $\Delta(t)$}

To compute $\Delta(t), C(t)$, and $R(t)$ we need information on $\mathrm{Im}\left[G(\omega) \right]$. Using Eqs.~\eqref{eq25},~\eqref{eq_re_gamma_Rubin_kneq_k1} and \eqref{eq_im_gamma_Rubin_kneq_k1} we get

\begin{align}\label{eq_Rubin_G}
&\mathrm{Im}\left[ G(\omega) \right] = \nn\\
& \frac{ \frac{k'^2}{k} \frac{1}{m\omega^3} \sqrt{\frac{1}{mk}}  \sqrt{1-\frac{m \omega ^2}{4 k}} \left( \left[\frac{k'}{k}\right]^2+ \frac{m \omega ^2}{k} \left[1-\frac{k'}{k}\right]  \right) } {\left[ \frac{M\omega^2}{k}\left[ 1-\frac{k'}{k} \right] +  \frac{k'^2}{k^2} \left[ \frac{M}{m} + \frac{1}{2} \right] -\frac{k'}{k} \right]^2 + \frac{k'^4}{m\omega^2 k^3} \left[ 1-\frac{m\omega^2}{4k} \right] } ;
\end{align}
for $|\omega| \leq 2\sqrt{k/m}$ and $0$ for $|\omega| > 2\sqrt{k/m}$.

The corresponding forms for the Drude bath is given by:
\begin{align}
\mathrm{Im}[G(\omega)] = \frac{\gamma_0 }{\omega \left[ M^2\omega^2 + (M\omega^2 \tau - \gamma_0)^2 \right]}.
\end{align}

Ohmic bath is obtained from the above expression by letting $\tau \rightarrow 0$. Most of the integrals of $\Delta(t)$ and $C(t)$ for the Rubin model are intractable analytically, but can be done numerically to extract some limiting behaviors:

Numerical results from the evaluation of the integral Eq.~\eqref{eq_delta_t} and a comparison with results from the corresponding Drude and Ohmic limits is shown in Figs.~(\ref{fig6}), (\ref{fig7}), (\ref{ShortTimeDelta}), (\ref{deltacomp}), and (\ref{deltacompLT}). 

Some of the interesting  observations can be summarized as follows:
\begin{enumerate}
\item  At long times we see a linear growth of $\Delta(t)$ with time, at all finite temperatures, as expected for a diffusive system.  
We notice that the integrand has an oscillatory factor $[1-\cos(\omega t)]$, so at large $t$ the major contribution to the integral comes from $\omega <<1/t$.
Hence, for any non-vanishing  $\beta$, we take $\coth\left(\frac{\beta\hbar\omega}{2}\right) \to \frac{2}{\beta\hbar\omega}$ to get (in the $t \to \infty$ limit):
\begin{align}
&\hspace{10pt} \Delta(t)  = \frac{4}{\pi\beta}\int_0^\infty d\omega \Big( \omega \mathrm{Im} [G(\omega)] \Big)_{\omega\rightarrow 0} \frac{1-\cos(\omega t)}{\omega^2}\nn\\
& \hspace{30pt} = \frac{2}{\beta\sqrt{mk}} t = 2Dt ~~ \label{eq41}\\
&\hspace{10pt} \mathrm{where}~~ D = \frac{1}{\beta\sqrt{mk}}=\frac{k_B T}{\gamma_0} \label{eq42}
\end{align}
can be identified  as the usual diffusion constant satisfying the Stokes-Einstein relation. In Fig.~(\ref{deltacomp}b) and (\ref{fig6}b), we verify that $\Delta(t)/t$ does converge to this limit at finite temperatures.
As can be seen from this figure, the long time asymptotics of $\Delta(t)$ and the diffusion constant are thus correctly obtained by both the Drude and Ohmic  limits. The Diffusion constant values are specified in Fig.~(\ref{deltacomp}b) for two different $\beta$ values and other parameter choices. Note that $D$ vanishes at zero temperature ($\beta = \infty$), which is also clear from the Fig.~(\ref{deltacomp}b).

\begin{figure*}
    \centering
    \includegraphics[width=\textwidth,height=5cm]{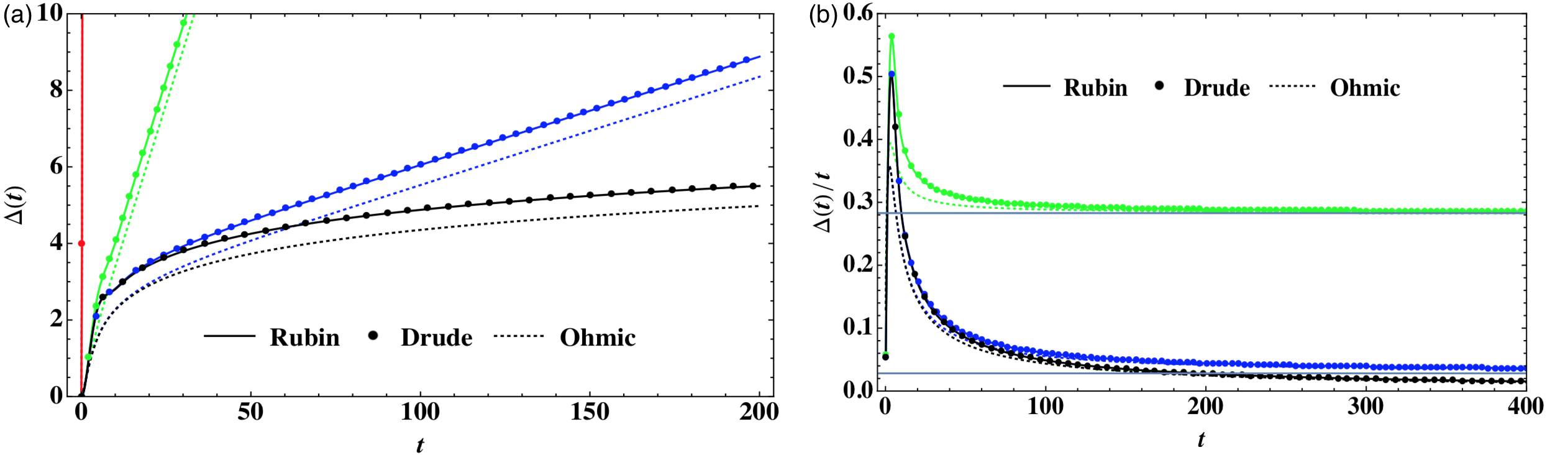}
   \caption{\textbf{Comparison of $\bm{\Delta(t)}$ between the Rubin, Drude and Ohmic models:}
(a)The $\beta$ values are $0.01, 10, 100$ and $\infty$ (from above). (b) For the $\beta$ values $10, 100, \infty$ (from above) we plot $\Delta(t)/t$. We verify the
asymptotic formulas presented in Eq.~\eqref{eq41} and \eqref{eq42} at finite temperatures.
At $\beta = 10$, the saturation value is $2D = 2/\beta\sqrt{km}=0.283$ and for $\beta =100$ it's $0.0283$ which match with the data.
These saturation values are indicated in the figure. Other parameters were taken as $M = 1, m = 0.1, k = 5, \bm{k' = 0.5}$.}
\label{deltacomp}
\end{figure*}

\begin{figure*}
    \centering
    \includegraphics[width=\textwidth,height=5cm]{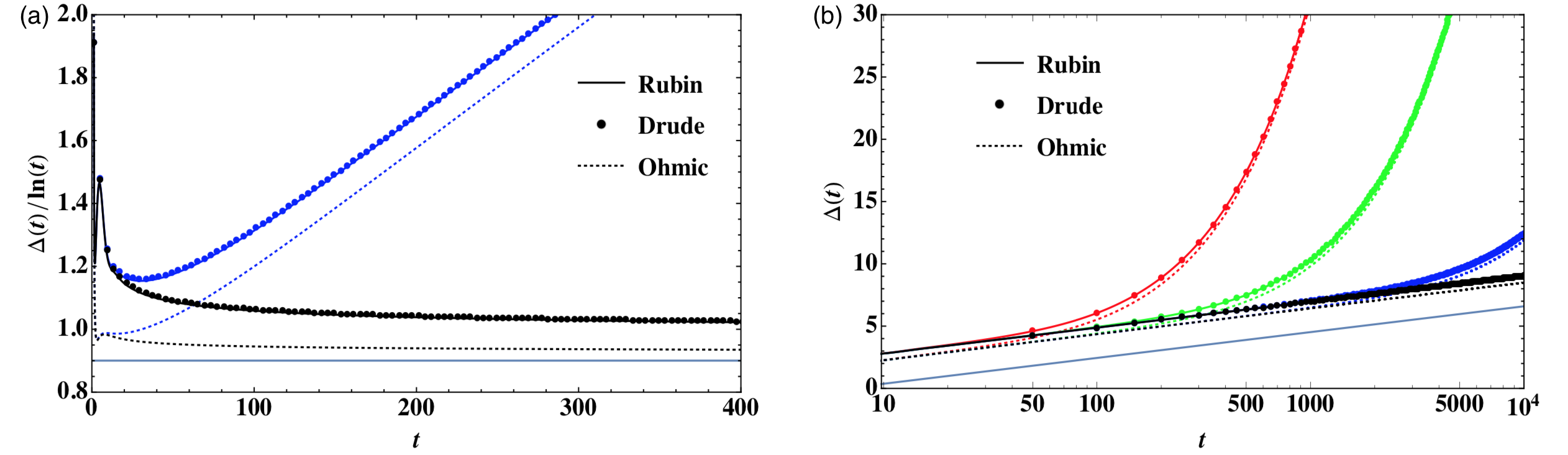}
   \caption{(a) Comparison of $\Delta(t)/\ln (t)$ between 
Rubin, Drude and Ohmic $\beta=10,\infty$ (from above) in linear scale. 
(b) $\Delta(t)$ in log-linear scale for $\beta = 100, 500, 5000, \infty$ (from above). 
This figure indicates the log behavior of $\Delta(t)$ for 
all three models with same pre-factors (slopes of the linear region) 
up to a time scale of the order of $\sim \beta \hbar$. 
Beyond this timescale $\Delta(t)$ behaves linearly in time 
which causes an exponential growth in log-linear scale. 
Other parameters were taken as $M = 1, m = 0.1, k = 5, \bm{k' = 0.5}$. 
We note that the Rubin and Drude models are matching well but the Ohmic is deviating. 
This is happening because $k$ is large but $k'$ is relatively small. 
However, as (b) suggests, the prefactors of $\ln(t)$ are same for these models and 
hence there is a slow convergence of the data at (a) for $\beta = \infty$. From Eq.~\eqref{eq_43} 
the prefactor of $\ln(t)$ is $2\hbar/\pi\gamma_0 = 0.9$ which has been indicated both in (a) and (b).}
\label{deltacompLT}
\end{figure*}

\item  At zero temperature $(\beta\rightarrow\infty)$ we find that $\Delta(t)$
has a slower logarithmic growth at large times.  
In this quantum regime we have  $\coth\left( \frac{\beta\hbar\omega}{2} \right) = 1$ and the integrals simplify. As before, we have to consider only the small $\omega$ contribution to the integral for the large time asymptotic behavior of $ \Delta(t) $
\begin{align}\label{eq_43}
\hspace{25pt}\Delta(t) 
& \simeq \frac{2\hbar}{\pi}\int_0^{2 \sqrt{\frac{k}{m}}} d\omega \Big( \omega \mathrm{Im} [G(\omega)] \Big)_{\omega\rightarrow 0} \frac{1-\cos(\omega t)}{\omega}\nn \\
& \simeq  \frac{2\hbar}{\pi \gamma_0} \ln (t)~.
\end{align} 
In Fig.~(\ref{deltacompLT}a) we verify this form and the value of the prefactor of $\ln (t)$. We see that Rubin, Drude and Ohmic models reproduce the logarithmic growth. The prefactors of $\ln(t)$ are same, which is evident from Fig.~(\ref{deltacompLT}b) and Fig.~(\ref{fig7}b), as the slopes of different models of the linear regime are same in log-linear scale. Note that there must be a timescale included in the argument of the log for dimensional constraints. The log behavior can be represented by, $\Delta(t) \sim A + B \ln(t)$, which implies that $\Delta(t)/\ln(t) \sim A/\ln(t) + B$. As $\ln(t)$ is a slowly varying (increasing) function of $t$, there is a slow convergence to the model independent pre-factor B of $\ln(t)$, as seen in Fig.~(\ref{deltacompLT}a) and Fig.~(\ref{fig7}a). The chosen parameters are mostly $M=1, m=0.1, k=5, k'=0.5$ or $4$ throughout the numerical data presented here for various $\beta$ values. For Drude and Ohmic baths, $\gamma_0$ and $\tau$ are also chosen correspondingly [see Eq.~\eqref{eq_20}].

\item At finite temperatures, the  cross-over from the quantum (logarithmic growth) to the classical (linear growth) takes place at the time scale $ t_{\rm qc} \sim \beta \hbar$.  We study this timescale in the Fig.~(\ref{deltacompLT}b) and Fig.~(\ref{fig7}b), by plotting the $\Delta(t)$ for various $\beta$ values keeping $\hbar=1$. In the log-linear scale, the log behavior of $\Delta(t)$ is represented by a linear regime which persists up to a timescale of the order of $\beta \hbar$. 

\begin{figure*}
    \centering
    \includegraphics[width=\textwidth,height=5cm]{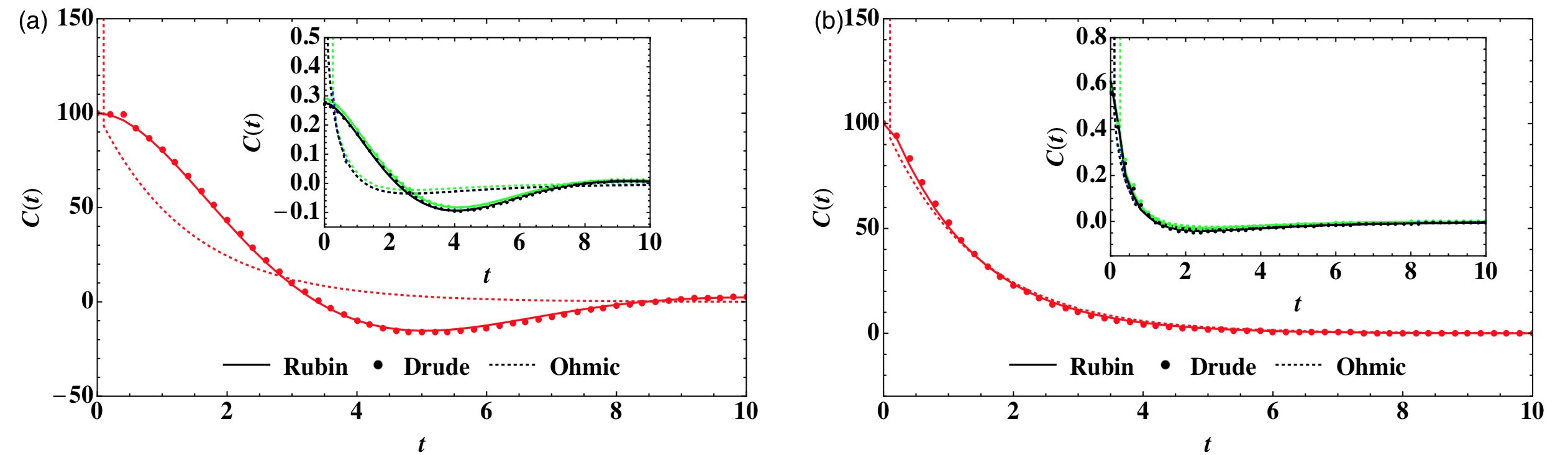}
   \caption{\textbf{Comparison of $\bm{C(t)}$ between the Rubin, Drude and Ohmic baths:} (a) $M = 1, m = 0.1, k = 5, \bm{k' = 0.5}$. 
The main figure shows $\beta=0.01$ and the inset shows $\beta= 10, 100, \infty$ (from above). (b) $M = 1, m = 0.1, k = 5, \bm{k' = 4}$. 
The main figure shows $\beta=0.01$ and the inset shows $\beta= 10, 100, \infty$ (from above). 
As in the case of $\Delta(t)$, there is a better agreement between the Rubin and Drude models than the Ohmic 
bath when $k$ value is sufficiently large but $k'$ is not. And all three models 
coincide when $k$ and $k'$ both are chosen to large. Near $t=0$, 
there is a log divergence, which is only present in the Ohmic case [see Eq.~\eqref{C_Ohmic_short_time}] and this deviation is always visible near $t=0$.}
\label{fig8}
\end{figure*}

\begin{figure*}
    \centering
    \includegraphics[width=\textwidth,height=5cm]{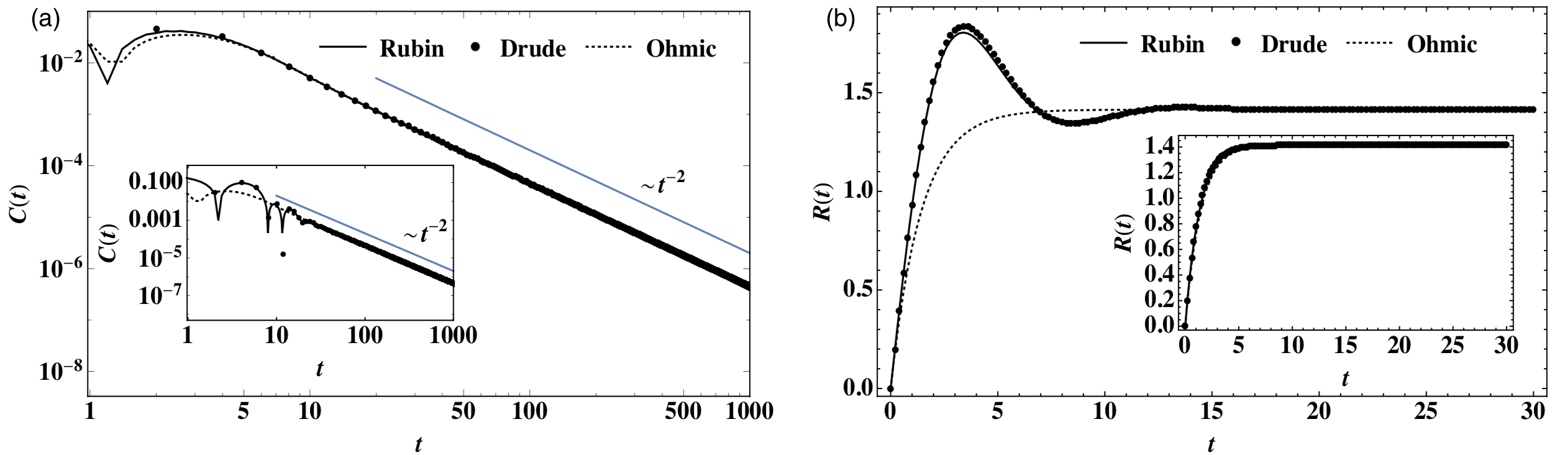}
   \caption{(a) $C(t)$ at large times for $M = 1, m = 0.1, k = 5, \bm{k' = 4}$ and 
$\beta = \infty$. The inset shows the same plot for $\bm{k'=0.5}$. $C(t)$ behaves as $1/t^2$ as discussed in the text. 
(b) Comparison of the response function $R(t)$ between the three models for $M = 1, m = 0.1, k = 5, \bm{k' = 0.5}$. The inset shows the same plot for 
$\bm{k'=4}$. $R(t)$ saturates to the value $R_\infty = 1/\gamma_0 = 1/\sqrt{km} = 1.414$ [see Eq.~\eqref{eq_48}].}
\label{fig10}
\end{figure*}

\item Finally we discuss the short time behavior. At  high temperatures, we approximate $\frac{1-\cos(\omega t)}{\omega^2}\sim t^2/2$ and $\coth\left(\frac{\beta\hbar\omega}{2}\right) \sim \frac{2}{\beta\hbar\omega}$, to obtain
\begin{align}
\hspace{12pt}\Delta(t)  \simeq \frac{2}{\pi\beta} t^2 \int_0^{2\sqrt{\frac{k}{m}}} d\omega~ \omega \mathrm{Im} [G(\omega)] \sim c \frac{k_B T}{M} t^2~,
\end{align}
where $c$ is a dimensionless constant. The ballistic growth can be simply understood as that of a thermal particle with $\la v^2 \ra=k_B T/M$.  
On the other hand at zero temperature, we get
\begin{equation}\label{eq_46}
\Delta(t) \simeq c' ~\frac{\hbar k^{1/2}}{M^{3/2}} t^2~,
\end{equation}
where $c' = [2 M^{3/2}/(k^{1/2} \pi)] \int_0^{2\sqrt{k/m}} d\omega~\omega^2 \mathrm{Im}[G(\omega)]$ is a dimensionless constant. The ballistic growth in this case roughly corresponds to a particle with velocity fluctuations determined by the zero point energy so that $\la v^2\ra = \hbar (k/M)^{1/2}/M$. 

As presented in the Fig.~(\ref{ShortTimeDelta}a), we see that in the high temperature limit, all three models show the $t^2$ behavior with same prefactor. This is consistent with the equipartition interpretation.

At zero temperature or any finite temperature, the Drude and Rubin models have the expected form of \eqref{eq_46} with same pre-factor while the Ohmic model has a logarithmic  correction given by: 
\begin{align}\label{Ohm_small_time}
\hspace{15pt}\Delta(t) \simeq -\frac{\hbar \gamma_0}{M^2 \pi} t^2 \ln \left( \gamma_0 t/M \right) + \mathcal{O}\big[t^4 \ln(t)\big].
\end{align}
The data is presented in Fig.~(\ref{ShortTimeDelta}b).

\end{enumerate}

\subsection{Form of  $C(t)$}
$C(t)$ is obtained from $\Delta(t)$ by taking two time derivatives  [Eq.~\eqref{eq_ct_def}]. Numerical data is presented in Fig.~(\ref{fig8}). Some important features are the following:
\begin{enumerate}
\item
One general feature is a damped oscillatory behavior in most of the parameter regimes. We can also see the agreement between the three models when both $k$ and $k'$ are chosen to be large. However there is a significant deviation of the Ohmic bath near $t=0$.

\item
In case of the Rubin and Drude model, at all temperatures, $\Delta(t)$ behaves as $\sim t^2$ near $t=0$ which gives a finite value of $C(0)$. Note that $C(0)=k_BT/M$ in classical regime and $C(0)=  \hbar (k/M)^{1/2}/M$ in quantum regime. $C(t) \sim C(0) + \mathcal{O}(t^2)$ at small times.

However, for Ohmic bath, from Eq.~\eqref{Ohm_small_time} we get,
\begin{align}\label{C_Ohmic_short_time}
C(t) \simeq - \frac{\gamma_0 \hbar}{\pi m^2} \ln \left( \gamma_0 t/M \right)  + \mathcal{O}\big[t^2 \ln(t)\big].
\end{align}
This log divergence near $t=0$ explains the deviation from other bath models shown in Fig.~(\ref{fig8}). Although the Eq.~\eqref{Ohm_small_time} was derived for low temperatures, this log divergence shows up at any finite temperature. In the classical limit, i.e. when $\beta \hbar = 0$, one gets an exponential decay of the velocity autocorrelation.

\item
In the previous section, we  obtained the leading order term for the $\Delta(t)$, which behaves as $\sim t$ in the large time limit at any finite temperature. If we take double derivatives naively, it does not lead to the correct leading order asymptote of $C(t)$. In a detailed calculation (to be published), we have shown that the correction to this linear behavior is $\sim e^{-c t}$ for the Drude and Ohmic models and $\sim \cos(\omega t)/t^{3/2}$ for the Rubin bath. Thus the large time behavior $(t >> \beta\hbar)$ of $C(t)$ is $\sim e^{-c t}$ for the Drude and Ohmic baths and $\sim \cos(t)/t^{3/2}$ for Rubin. At zero temperature or $t<<\beta\hbar$ the leading order behaviors are $\sim 1/t^2$ for all three models.

\end{enumerate}

\subsection{Form of  $R(t)$}
Using Eq.~\eqref{eq_R} we obtain,
\begin{align}\label{eq_48}
R(t) = \left[ 1 - \exp(-\gamma_0 t/m) \right]/\gamma_0
\end{align}
for the Ohmic model. For Drude model, the integrals can also be evaluated exactly and $R(t)$ takes similar functional form. The general feature that $R(t)$ increases initially and then saturates to a value is present in all models and parameter regimes. This behavior physically describes the fact that if we perturb the Brownian particle, it will initially have a directional displacement before it becomes completely random. For the Rubin bath the integrals are intractable. Data from numerical integration for all three modes are shown in Fig.~(\ref{fig10}).\\

\paragraph*{Numerical details:} To perform the integrals numerically \textsc{Mathematica} has been used extensively, especially the NIntegrate command. To obtain the analytical and asymptotic formulas, doing the summations, etc., the commands like Integrate, AsymptoticIntegrate, Series, FullSimplify, etc. of \textsc{Mathematica} have been particularly used.\\

\section{Summary and Discussion} \label{Summary}
In this paper we study in detail the well-known Rubin bath model, which consists of a one-dimensional semi-infinite harmonic chain with the boundary bath particle coupled to a test particle, which is then shown to effectively execute Brownian motion.

We point out two interesting and important limits of the Rubin model: (i) the Drude model which is obtained in the infinite bath bandwidth limit of the Rubin model and (ii) the Ohmic model which, in addition to an infinte bath bandwidth, also needs the limit of infinite system-bath coupling. For the Rubin model and the special limiting cases, we analyse in detail the temporal dependence of the mean square displacement, the velocity autocorrelation function and the response function. In addition, we studied the crossover behaviour of the 
dissipation kernel $\gamma(t)$  from an exponentially decaying behaviour at short times to an oscillatory power law ($\sim t^{-3/2}$) decaying behaviour at larger times.

Taking the special limits of either the Drude and Ohmic baths is useful since the bath kernels are much simpler and the mathematical analysis becomes considerably simpler. In real physical situations one might have large but finite bath bandwidths and system-bath couplings. One important question in these situations is as to how  closely physical properties are reproduced when we ignore the fact that the original Rubin bath kernel has long time power-law tails. Our numerical results show that many properties are indeed accurately reproduced by the approximate models. For example, we show that even though the Rubin bath  has a memory kernel with power law tails, which is completely different from the exponential decay for the Debye model, this does not affect the asymptotic form of the MSD. In particular, we discussed the quantum to classical crossover time scales where the mean square displacement changes from a logarithmic to a linear time  dependence. The analysis presented in this work provides a microscopic justification for the choice of the position response function used in a recent analysis~\cite{urbashi2017} of quantum Brownian motion based on linear response theory as the starting point.

We have shown that the Ohmic limit, when the dissipation kernel becomes a $\delta$-function, is obtained in the limit of a continuum string and surprisingly, for strong coupling. This is unlike the weak coupling limits usually discussed in the derivation of quantum master equations in the literature \cite{carmichael2013,breuer2002}. An interesting observation is that at any finite temperature the noise correlations always have a finite correlation time even in the Ohmic limit when the dissipation kernel becomes a $\delta$-function. Thus a quantum bath is never truly Markovian. However at high temperatures one can make the approximation  $\coth (\beta \hbar \omega/2) \to 2/(\beta \hbar \omega)$  in the noise correlations and then get the Markovian limit. Thus our study shows the precise conditions under which the Markovian approximation is valid. We note that the microscopic derivation of quantum master equations typically starts with exactly the same system-bath setup as the one used in the derivation of the quantum Langevin equation.  There, the Born-Markov approximation leads to the Redfield equation and further approximations lead to the Lindblad equation which is Markovian. The precise conditions for the validity of the Born-Markov approximation are however subtle and not clearly understood \cite{carmichael2013,breuer2002,purkayastha2016} and we believe that our work, with very explicit results, could provide insights on this issue.

\begin{acknowledgements}
We acknowledge support of the Department of Atomic Energy, Government of India, under project no.12-R$\&$D-TFR-5.10-1100. IS acknowledges the S.N.Bhatt Memorial fellowship 2017 at ICTS.
\end{acknowledgements}

\bibliography{references}

\end{document}